\documentclass[sigconf, noacm, dvipsnames]{acmart}

\AtBeginDocument{%
  \providecommand\BibTeX{{%
    \normalfont B\kern-0.5em{\scshape i\kern-0.25em b}\kern-0.8em\TeX}}}

\usepackage{graphicx}
\usepackage{textcomp}
\usepackage{booktabs}
\usepackage{multirow}

\newcommand\ignore[1]{ }

\usepackage{enumitem} 

\usepackage{xcolor}
\usepackage{multirow}

\definecolor{airforceblue}{rgb}{0.36, 0.54, 0.66}
\definecolor{dodgerblue}{rgb}{0.12, 0.56, 1.0}
\definecolor{brandeisblue}{rgb}{0.0, 0.44, 1.0}
\definecolor{brickred}{rgb}{0.8, 0.25, 0.33}
\definecolor{eggplant}{rgb}{0.38, 0.25, 0.32}
\definecolor{purple}{rgb}{0.44, 0.16, 0.55}
\definecolor{ddgreen}{rgb}{0.00, 0.50, 0.00}

\newcommand{\juan}[1]{\textcolor{black}{#1}}

\definecolor{dgreen}{rgb}{0.00, 0.75, 0.00}
\newcommand{\juang}[1]{\textcolor{black}{#1}}
\newcommand{\juangg}[1]{\textcolor{black}{#1}}
\newcommand{\juanggg}[1]{\textcolor{black}{#1}}
\newcommand{\juanggr}[1]{\textcolor{black}{#1}}
\newcommand{\juanggrr}[1]{\textcolor{black}{#1}}

\usepackage{tikz}
\usetikzlibrary{shapes.geometric, arrows, positioning, shadows}

\pgfdeclarelayer{background}
\pgfsetlayers{background,main}

\tikzstyle{texte} = [above]

\tikzstyle{stop} = [rectangle, rounded corners, minimum width=3cm, minimum height=1cm,text centered, draw=black, fill=violet!30, drop shadow]
\tikzstyle{io} = [trapezium, trapezium left angle=70, trapezium right angle=110, text centered, text width=5cm, draw=black, fill=blue!30, drop shadow]
\tikzstyle{processcpu} = [rectangle, text centered, text width=5cm, draw=black, fill=orange!30, drop shadow]
\tikzstyle{processdpu} = [rectangle, text centered, text width=5cm, draw=black, fill=red!30, drop shadow]
\tikzstyle{decision} = [diamond, aspect=2, text centered, draw=black, fill=green!30, drop shadow]
\tikzstyle{arrow} = [thick,->,>=stealth]

\usepackage{algorithm}
\usepackage{algpseudocode}
\usepackage{stmaryrd}
\usepackage{amsmath}
\usepackage{listings}
\usepackage{subcaption}

\definecolor{mygreen}{rgb}{0,0.6,0}
\definecolor{mygray}{rgb}{0.5,0.5,0.5}
\definecolor{mymauve}{rgb}{0.58,0,0.82}

\setcopyright{none}
\renewcommand\footnotetextcopyrightpermission[1]{}

\newcommand{\libname}{{TransPimLib}}

\newcommand{\tklbl}[0]{Key Takeaway}
\newcounter{take}
\newcommand\takeaway[1]{%
   \stepcounter{take}
   \noindent
   \colorbox{gray!20}{\textbf{\tklbl{} \thetake.}} \emph{#1}}

\begin{document}

\title{\libname: A Library for Efficient Transcendental Functions \\ on Processing-in-Memory Systems}

\author{
 {%
     Maurus Item\quad\hspace{0.5cm}
     Juan Gómez-Luna\quad\hspace{0.5cm} 
     Yuxin Guo\quad 
 }
}
\author{
 {
     Geraldo F. Oliveira\quad\hspace{0.5cm}
     Mohammad Sadrosadati\quad\hspace{0.5cm}
     Onur Mutlu
 }
}


\affiliation{
\institution{
      \vspace{5pt}
      ETH Zürich \quad
  }
}

\pagestyle{plain}

\begin{abstract}
Processing-in-memory (PIM) promises to alleviate the data movement bottleneck in modern computing systems. 
However, current real-world PIM systems have the inherent disadvantage that their hardware is more constrained than in conventional processors (CPU, GPU), due to the difficulty \juang{and cost} of building processing elements near or inside the memory. 
As a result, general-purpose PIM architectures support fairly limited instruction sets and struggle to execute complex operations such as transcendental functions and other hard-to-calculate operations (e.g., square root). 
These operations are particularly important for some modern workloads, e.g., activation functions in machine learning applications.

In order to provide support for transcendental (and other hard-to-calculate) functions in general-purpose PIM systems, we present \emph{\libname}, a library that provides CORDIC-based and LUT-based methods for trigonometric functions, hyperbolic functions, exponentiation, logarithm, square root, etc. 
We develop an implementation of \libname~for the UPMEM PIM architecture and perform a thorough evaluation of \libname's methods in terms of performance and \juangg{accuracy}, using microbenchmarks and \juan{three} full workloads (Blackscholes, \juan{Sigmoid}, Softmax).
\juang{We open-source all our code and datasets at~\url{https://github.com/CMU-SAFARI/transpimlib}.}

\end{abstract}


\maketitle

\section{Introduction}


\juangg{Processor} performance increasing more rapidly than memory performance \juangg{for decades has caused} a wide gap between processing units and memory units in terms of latency and energy consumption. 
As a result, access to data has become a major bottleneck in current computing systems~\cite{mutlu2020modern, kestor.iiswc2013, pandiyan2014, boroumand2018, deoliveira2021IEEE, boroumand2021google}. 
Processing-in-memory (PIM) is a \juangg{promising} solution to this data movement bottleneck. PIM consists of equipping memory with computing capabilities, either with small functional units \emph{near} the memory arrays or by \emph{using} the analog \juangg{operational properties} of memory cells themselves~\cite{mutlu2020modern, seshadri.micro17, hajinazarsimdram}. PIM provides access to data at significantly higher bandwidth, lower latency, and lower energy consumption than conventional \juangg{compute-centric} processors (e.g., \juangg{CPUs, GPUs}). 

Explored for more than 50 years~\cite{Kautz1969,stone1970logic}, PIM's materialization into real products was delayed \juangg{in part} by fundamentally different requirements of fabrication of memory units and processing units~\cite{devaux2019, weber2005current, peng2015design,yuffe2011, christy2020, singh2017}. 
Only now we are witnessing the arrival of the first PIM commercial products and prototypes. 
UPMEM~\cite{upmem}, for example, introduced the first general-purpose commercial PIM architecture~\cite{upmem,upmem2018,gomezluna2021benchmarking, gomezluna2022ieeeaccess, gomezluna2021cut}, which integrates \juangg{a small in-order core next to each DRAM bank in a DRAM chip}. 
HBM-PIM~\cite{kwon202125, lee2021hardware} and Acceleration DIMM (AxDIMM)~\cite{ke2021near} are Samsung's \juangg{real PIM prototypes}. 
HBM-PIM features \juangg{a SIMD unit that supports multiply-add (MAD) and multiply-accumulate (MAC) operations between every two banks in HBM~\cite{jedec.hbm.spec, lee.taco16} layers. HBM-PIM} is designed to accelerate neural network inference. 
AxDIMM is a near-rank solution that places an FPGA fabric on a DDRx module to accelerate specific workloads (e.g., recommendation inference). 
Accelerator-in-Memory (AiM)~\cite{lee2022isscc} is a GDDR6-based PIM architecture from SK Hynix with specialized units for multiply-accumulate and activation functions for deep learning. 
HB-PNM~\cite{niu2022isscc} is a 3D-stacked-based PIM architecture from Alibaba, which stacks together a layer of LPDDR4 memory and a logic layer with specialized accelerators for {recommendation} systems. 
Among other characteristics, these PIM systems have in common that (1) they place PIM processing elements \emph{near} memory, and (2) these PIM processing elements are relatively simple and natively support  \emph{limited} instruction sets (e.g., integer arithmetic in UPMEM, 16-bit floating-point MAC/MAD in HBM-PIM or AiM). 

Given that instruction sets of current real-world PIM architectures are limited, some complex operations should be emulated by the runtime library (e.g., integer multiplication/division and floating-point arithmetic in UPMEM PIM~\cite{upmem-guide, gomezluna2022ieeeaccess}) or are not even supported. 
This is usually \juangg{true for} \emph{transcendental functions} (e.g., trigonometric, hyperbolic, exponentiation, logarithm) and other hard-to-calculate functions (e.g., square root) in PIM architectures. 
These functions are used in a variety of important applications, such as activation functions in machine learning applications~\cite{hosmer2013applied, han1995influence}, finite element methods~\cite{bathe2007finite}, ray tracing~\cite{glassner1989introduction}, and option pricing in the stock market~\cite{macbeth1979empirical}. 
\juanggr{Prior work~\cite{memoization1, suresh2016intercepting} has identified the presence of transcendental functions that are executed thousands of times in several workloads from various benchmark suites (e.g., SPEC CPU2006~\cite{spec2006}, SPLASH-2~\cite{woo_isca1995}).} 
Many of these applications suffer from the data movement bottleneck in conventional processors and, thus, can potentially benefit from PIM. 
For example, a recent PIM benchmarking study~\cite{deoliveira2021IEEE} identified seven memory-bound applications that use transcendental functions. 

To cope with the need for efficient support for transcendental functions and other hard-to-calculate functions in PIM architectures, we analyze possible alternatives to compute these functions. 
In current computing systems with PIM capabilities~\cite{upmem, upmem-guide, gomezluna2022ieeeaccess, lee2022isscc, kwon202125, lee2021hardware, devaux2019, gomezluna2021benchmarking, gomezluna2021cut}, there is a host processor (e.g., CPU, GPU) that offloads memory-bound computations to the PIM processing elements. Offloading may require moving data from the standard main memory to the PIM-enabled memory. 
We identify three possible alternatives to compute transcendental functions in such systems. 

\begin{figure}[h]
\centering
\includegraphics[width=1.0\linewidth]{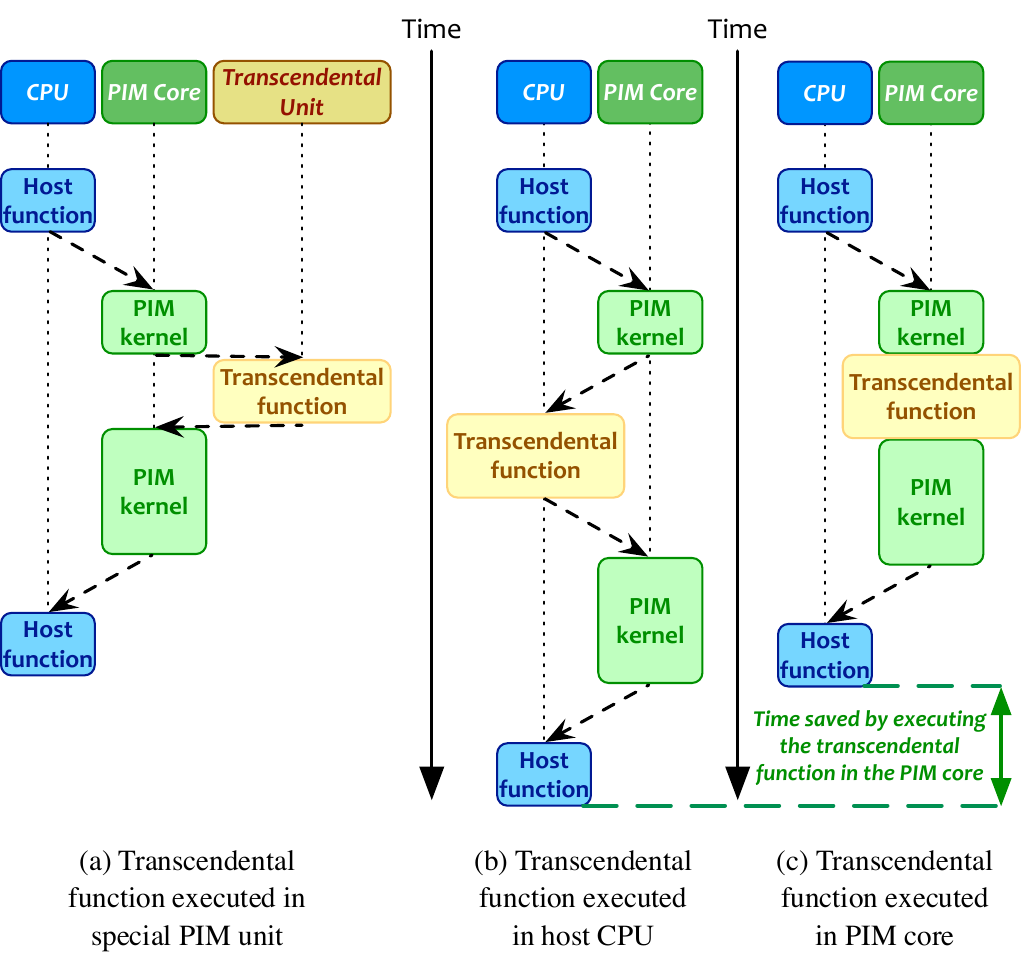}
\caption{\juangg{Three options for calculation} of transcendental functions in a PIM system.}
\label{fig:transcendental}
\end{figure}

First, the PIM hardware can integrate a special hardware unit for transcendental functions. As Figure~\ref{fig:transcendental}(a) represents, a PIM core can invoke a ``transcendental unit'', which executes the desired function. 
Such special units are common in conventional processors, e.g., specialized co-processors in CPUs~\cite{intel2007} and special function units in GPUs~\cite{cuda-guide}. 
To our knowledge, the only real PIM architecture that provides this kind of support is AiM~\cite{lee2022isscc}, which employs lookup tables (LUTs) and a hardware interpolation unit for different activation functions. 
Such support may \emph{not} 
(1)~be affordable for all PIM designs (given the area cost of additional hardware units~\cite{doubleprecisioncordic, freescale_coprocessor_manual}), 
(2)~provide a flexible implementation \juangg{(i.e., with support for different methods which can better suit different functions)}, and 
(3)~provide \juangg{flexible} control on the desired precision. 

Second, we can compute transcendental functions on the host processor, as shown in Figure~\ref{fig:transcendental}(b). 
This approach has two drawbacks: (1) programmers of computing systems with PIM capabilities need to \juangg{properly} partition the \juangg{applications}, and (2) data needs to move back-and-forth between PIM cores and the host processor, which hampers the potential benefit from PIM. 

Third, we can implement fast and self-contained calculation methods for transcendental functions using existing instructions (native or emulated) \juangg{in the PIM core}, as Figure~\ref{fig:transcendental}(c) shows. 
While polynomial approximations \cite{Clenshaw1954PolynomialAT, transc_taylor_minmax, transc_taylor} are frequently used \juangg{such} methods, there is no study of other methods, such as CORDIC~\cite{volder1959cordic}, or a comprehensive exploration of LUT-based methods~\cite{lut1, lut2, lut3, lut4}. 

\juangg{Our} \textbf{goal} is to explore different methods for calculating transcendental and other hard-to-calculate functions in PIM systems. 
We develop our methods for the UPMEM PIM architecture~\cite{devaux2019, gomezluna2021benchmarking, gomezluna2022ieeeaccess, gomezluna2021cut}, the first publicly-available PIM architecture, which consists of general-purpose in-order cores placed near DRAM banks. 
As a result of our study, we present \emph{\libname}, \juangg{an open-source} library of transcendental and other hard-to-calculate functions for PIM. 
\libname~implements CORDIC-based methods, LUT-based methods, and combinations and variations of them (\juangg{with and without} interpolation). In total, \libname~uses eight different methods for trigonometric functions (sine, cosine, tangent), hyperbolic functions (sinh, cosh, tanh), exponentiation, logarithm, square root, and \juangg{GELU (Gaussian Error Linear Unit)~\cite{gelu}}.\footnote{\juangg{We open source \libname~to facilitate reproducibility and future research~\cite{transpimlib2023repo}.}}

We compare all \juangg{of} \libname's methods in terms of \juangg{accuracy}, execution cycles in the PIM cores, setup time in the host CPU, and memory consumption. 
\libname's methods provide \juangg{accuracy} of up to $10^{-9}$ root-mean-square absolute error (RMSE). 
LUT-based methods, in particular our \emph{LDEXP-based Fuzzy Lookup Table} (\emph{L-LUT}) method, demonstrate the best tradeoff \juangg{between} performance and \juangg{accuracy}. The non-interpolated L-LUT method requires no multiplication or other complex operations while achieving RMSE=$10^{-7}$. The interpolated L-LUT method has a \juangg{accuracy} of RMSE=$10^{-9}$ at the expense of just one multiplication. 

We evaluate \libname's functions for three full workloads: Blackscholes~\cite{macbeth1979empirical}, Sigmoid~\cite{han1995influence}, and Softmax~\cite{alpaydin2020}. 
The fastest PIM version of Blackscholes outperforms a 32-thread CPU baseline by 62\%. 
The PIM versions of Sigmoid and Softmax, which are typically used as activation functions in neural networks, provide competitive performance to their 32-thread CPU counterparts and prove that \libname~can \juangg{reduce} data movement between PIM cores and the host CPU (\juangg{as it is needed in Figure~\ref{fig:transcendental}(b), where transcendental functions run in the host CPU}). 

Our main contributions are as follows:
\begin{itemize}
\item We present \libname, the first library of transcendental and other hard-to-calculate functions for PIM architectures. \libname~contains CORDIC-based and LUT-based methods, and combinations and variations of them. 
\item We propose new LUT-based methods, called \juangg{\emph{L-LUT}}, \emph{D-LUT}, and \emph{DL-LUT} (Section~\ref{sec:lib-table}), which demonstrate good suitability for general-purpose PIM architectures with limited instruction sets, such as \juangg{the UPMEM PIM system}. 
\item We evaluate \libname's methods in terms of \juangg{accuracy}, execution cycles in the PIM cores, setup time in the host CPU, and memory consumption. We also evaluate \libname's functions for three real workloads (Blackscholes, Sigmoid, Softmax). 
\item We open-source \libname, as well as all codes and datasets used for evaluation, \juang{in our GitHub repository~\cite{transpimlib2023repo}}.
\end{itemize}

\section{Background}
\label{sec:background}

This section provides the necessary background on current real-world processing-in-memory systems (Section~\ref{sec:pim}), and on transcendental functions and calculation methods (Section~\ref{sec:transcendental}).

\subsection{Processing-in-Memory (PIM)}
\label{sec:pim}

Processing-in-memory {(PIM)} is a computing paradigm that advocates for memory-centric computing systems, where \juanggg{memory becomes an active system component with computing capabilities. 
These capabilities can be either (1) small processing elements (general-purpose cores and/or accelerators) placed \emph{near} the memory arrays (e.g.,~\cite{deoliveira2021,boroumand.asplos18,boroumand2019conda,ahn.tesseract.isca15,syncron,singh2019napel,zhu2013accelerating, DBLP:conf/isca/AkinFH15, DBLP:conf/sigmod/BabarinsaI15, kim.bmc18, cali2020genasm,fernandez2020natsa,impica,singh2020nero,ahn.pei.isca15,nai2017graphpim,hadidi2017cairo,zhang.hpdc14, pattnaik.pact16, hsieh.isca16,kim.sc17,DBLP:conf/hpca/GaoK16, guo2014wondp, asghari-moghaddam.micro16}, or 
(2) mechanisms that compute by \emph{using} the analog operational properties of memory components (e.g., by simultaneously activating multiple memory cells~\cite{aga.hpca17,eckert2018neural,fujiki2019duality,kang.icassp14,seshadri2020indram,seshadri.micro17,seshadri2013rowclone,kim.hpca18,kim.hpca19,gao2020computedram,chang.hpca16,li.micro17,hajinazarsimdram,wang2020figaro,olgun2021quactrng,li.dac16,angizi2018pima,angizi2018cmp,angizi2019dna,levy.microelec14,kvatinsky.tcasii14,shafiee2016isaac,kvatinsky.iccd11,kvatinsky.tvlsi14,gaillardon2016plim,bhattacharjee2017revamp,hamdioui2015memristor,xie2015fast,hamdioui2017myth,yu2018memristive,puma-asplos2019, ankit2020panther,chi2016prime,ambrosi2018hardware,bruel2017generalize,huang2021mixed,zheng2016tcam,xi2020memory}). 
First} proposed more than 50 years ago~\cite{Kautz1969,stone1970logic}, processing-in-memory is a \juanggg{compelling} solution to alleviate the \emph{data movement bottleneck}~\cite{mutlu2019,mutlu2020modern,mutlu2021intelligent}, \juanggg{which is} caused by the need for moving data between memory units and compute units in processor-centric systems. \juanggg{Such bottleneck has only become worse over the years due to} the faster development of processor performance over memory performance.

PIM architectures are becoming a reality, with the commercialization of the UPMEM PIM architecture~\cite{upmem,upmem2018,gomezluna2021benchmarking}, and the announcement of HBM-PIM~\cite{kwon202125,lee2021hardware}, AxDIMM~\cite{ke2021near}, AiM~\cite{lee2022isscc}, and HB-PNM~\cite{niu2022isscc} (all four prototyped and evaluated in real systems). 
\juanggg{UPMEM places a small general-purpose in-order core (called \emph{DPU}) near each memory bank of a DDR4 DRAM chip. 
HBM-PIM features a SIMD unit (called \emph{PCU}) between every two banks in the memory layers of an HBM stack~\cite{jedec.hbm.spec}. HBM-PIM is designed for machine learning inference. Thus, its SIMD units execute only a reduced set of instructions (i.e., 16-bit floating-point multiplication and addition). 
AxDIMM~\cite{ke2021near} is a DIMM-based solution with an FPGA inside the buffer chip of the DIMM. The FPGA can accelerate memory-bound workloads, such as recommendation inference~\cite{ke2021near} or database operations~\cite{lee2022improving}. 
AiM~\cite{lee2022isscc} is a GDDR6-based PIM architecture with a near-bank processing unit (called \emph{PU}) that executes multiply-and-accumulate operations and activation functions. Same as HBM-PIM, AiM targets machine learning workloads. 
HB-PNM~\cite{niu2022isscc} is a 3D-stacked based PIM solution for recommendation systems. HB-PNM stacks one layer of LPDDR4 DRAM~\cite{jedec2021lpddr4} on one logic layer connected through hybrid bonding (HB) technology~\cite{fujun2020stacked}. The logic layer embeds two types of specialized engines for \emph{matching} and \emph{ranking}, which are the memory-bound steps of the evaluated recommendation system.}

\begin{figure}[h]
\centering
\includegraphics[width=1.0\linewidth]{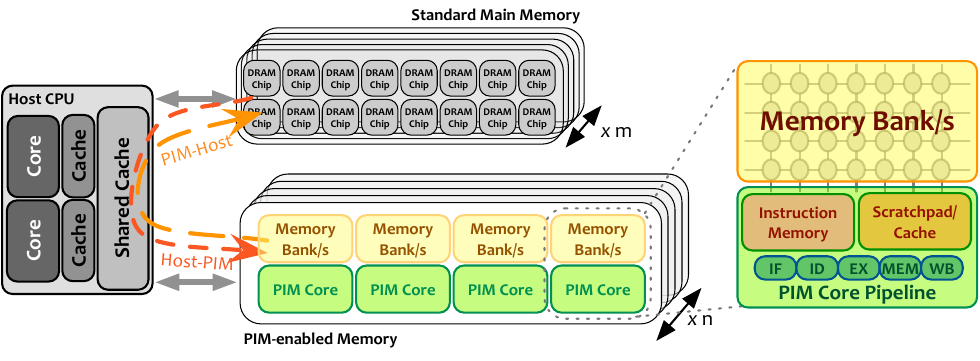}
\caption{\juanggg{Example} state-of-the-art processing-in-memory system with general-purpose PIM cores. The host CPU has access to $m$ standard memory modules and $n$ PIM-enabled memory modules.}
\label{fig:scheme}
\end{figure}

These five real-world PIM systems have some important common characteristics, as depicted in Figure~\ref{fig:scheme}. 
First, there is a host processor (CPU or GPU), typically with a deep cache hierarchy, which has access to (1) standard main memory, and (2) PIM-enabled memory (i.e., UPMEM DIMMs, HBM-PIM stacks, AxDIMM DIMMs, AiM GDDR6, HB-PNM LPDDR4). 
Second, the PIM-enabled memory chip contains multiple PIM processing elements (PIM PEs), which have access to memory (either memory banks or ranks) with \juanggg{much} higher bandwidth and lower latency than the host processor. 
Third, the {PIM} processing elements (either {general-purpose} cores, SIMD {units, FPGAs, or specialized processors}) run at {only} a few hundred megahertz, and have a small number of registers and 
{relatively small} (or no) cache or scratchpad memory. 
Fourth, PIM PEs may not be able to communicate directly {with each other} (e.g., UPMEM DPUs, HBM-PIM PCUs or AiM PUs in different chips), 
{and} communication {between them} happens via the host processor. 
Figure~\ref{fig:scheme} shows a high-level view of such a state-of-the-art processing-in-memory system. 

In \juanggg{this} work, we use the UPMEM PIM architecture~\cite{devaux2019, upmem, upmem2018, upmem-sdk, gomezluna2021benchmarking, upmem-guide}, the first PIM architecture to be commercialized in real hardware. 
The UPMEM PIM architecture uses conventional 2D DRAM arrays and combines them with general-purpose processing cores, called \emph{DRAM Processing Units} (\emph{DPUs}), on the same chip. 
There are 8 DPUs and 8 DRAM banks per chip, and 16 chips per DIMM (8 chips/rank). 
DPUs are relatively deeply pipelined and fine-grained multithreaded~\cite{burtonsmith1978,smith1982architecture,thornton1970}. 
DPUs run software threads, called \emph{tasklets}, which are programmed in \emph{Single Program Multiple Data} (\emph{SPMD}) manner. 

DPUs have a 32-bit RISC-style general-purpose instruction set~\cite{upmem-guide}. They feature native support for 32-bit integer, but some complex operations (e.g., 32-bit integer multiplication/division) and {floating-point} operations are emulated~\cite{gomezluna2021benchmarking}. 

Each DPU has exclusive access to its own (1) 64-MB DRAM bank, called \emph{Main RAM} (\emph{MRAM}), (2) 24-KB instruction memory, called \emph{Instruction RAM} (\emph{IRAM}), and (3) 64-KB scratchpad memory, called \emph{Working RAM} (\emph{WRAM}). 
The host CPU can access the MRAM banks for copying input data (from main memory to MRAM) and retrieving results (from MRAM to main memory). These CPU-DPU/DPU-CPU transfers can be performed in parallel (i.e., concurrently across multiple MRAM banks), if the size of the buffers transferred from/to all MRAM banks is the same. Otherwise, the data transfers should be performed serially. Since there is no direct communication channel between DPUs, all inter-DPU communication takes place through the host CPU by using DPU-CPU and CPU-DPU data transfers. 
\juanggg{We refer the reader to our prior work~\cite{gomezluna2021cut, gomezluna2021benchmarking, gomezluna2022ieeeaccess} for a comprehensive introduction to and analysis of the UPMEM PIM system.}

Throughout this paper, we use generic terminology, since our implementation strategies are applicable to PIM systems like the generic one described in Figure~\ref{fig:scheme}, and not exclusive \juanggg{to} the UPMEM PIM architecture. 
Thus, we use the terms \emph{PIM core}, \emph{PIM thread}, \emph{DRAM bank}, \emph{scratchpad}, and \emph{Host-PIM/PIM-Host transfer}, which correspond to DPU, tasklet, MRAM bank, WRAM, and CPU-DPU/DPU-CPU transfer in UPMEM's terminology~\cite{upmem-guide}.

\subsection{Transcendental Functions}
\label{sec:transcendental}
\emph{Transcendental functions}~\cite{encyc_math_transc, transc_original} are functions that do \emph{not} satisfy a polynomial equation. 
As a result, they cannot be exactly expressed with a finite \juanggg{number} of algebraic operations (e.g., addition, subtraction, multiplication, division, power, root). 
Commonly-used transcendental functions are trigonometric functions (e.g, sine, cosine), exponential functions, and \juanggg{logarithm functions}. 
In this work, we also target other functions \juanggg{that} are not transcendental, but hard to calculate (e.g., square root). 

There are various methods to calculate transcendental functions, such as \emph{Taylor approximation}~\cite{transc_taylor, transc_taylor_minmax}, \emph{minimax polynomials}~\cite{transc_taylor_minmax}, \emph{CORDIC}~\cite{volder1959cordic, cordicprecision, 50ycordic}, and \emph{table-based methods}~\cite{lut1, lut2, lut3, lut4}. 
In this work, we focus on (1) CORDIC~\cite{volder1959cordic} and (2) \emph{fuzzy lookup tables}~\cite{lut1, lut2, lut3, lut4}, since these methods have low usage of floating-point multiplication (more expensive than addition/subtraction in current general-purpose PIM architectures~\cite{upmem-sdk, gomezluna2022ieeeaccess}). 
We introduce both methods in Sections~\ref{sec:cordic} and~\ref{sec:fuzzy}. In Section~\ref{sec:range}, we \juanggg{describe} how we extend the input range of these methods. 

\subsubsection{CORDIC}
\label{sec:cordic}
CORDIC \cite{volder1959cordic} is an iterative method that uses only bit-shifts, additions, and table lookups. 
The maximum error shrinks exponentially with the number of iterations~\cite{cordicprecision}. 
We use CORDIC in \emph{rotation mode}~\cite{volder1959cordic}, which computes a function value (e.g., sine, cosine) for a given input (an angle $\theta$). 
CORDIC starts with an angle $\theta_0 = \theta$ and a vector $v_0 = \begin{bmatrix}  1\\ 0 \end{bmatrix}$. 
In each iteration ($i > 0$), we update $\theta_i$ and rotate the vector by multiplying it with a $2 \times 2$ matrix $M_i$: $v_{i+1} = k_i \cdot M_i \cdot v_i$. 
$k_i$ is a stretching factor. 
$M_i$ represents a rotation by an angle $\phi_i$, which decreases in each iteration. The values of $\phi_i$ can be precalculated and kept in a table. 
After $n$ iterations, the output vector $v_n$ contains function values for $\theta$. 
Table~\ref{tab:cordicmethods} shows common rotation matrices, angles, stretching factors, and functions they are used for.

\begin{table}[h]
\caption{CORDIC's Rotation Matrices, Angles, and Stretching Factors}
\label{tab:cordicmethods}
\resizebox{\columnwidth}{!}{
\begin{tabular}{|l|ccc|l|}
\hline
Matrix & $M_i$   & $\phi_i$  & $k_i$  & Functions\\ 
\hline
\hline

Circular    
& $\begin{bmatrix} 1 & \mp 2^{-i}\\ \pm 2^{-i} & 1 \end{bmatrix}$
& $\pm arctan(2^{-i})$
& $\sqrt{1 + 2^{-2i}}$ 
& $sin, cos, tan, arctan$ \\
\hline

Hyperbolic         
& $\begin{bmatrix} 1 & \pm 2^{-i}\\ \pm 2^{-i} & 1 \end{bmatrix} $
& $\pm atanh(2^{-i}) $
& $\frac{2^{-i-1} + 1}{2^{-i-1}}$
& $sinh, cosh, tanh, exp, log, sqrt, atanh$ \\
\hline

Linear
& $\begin{bmatrix} 1 & 0 \\ \pm 2^{-i} & 1 \end{bmatrix} $
& $\pm 2^{-i}$
& 1
&\juanggg{multiplication}, division \\
\hline
\end{tabular}
}
\end{table}

\subsubsection{Fuzzy Lookup Tables}
\label{sec:fuzzy}
Lookup tables are widely used to generate a variety of functions~\cite{lut1, lut2, lut3, lut4}. 
Since lookup tables are limited in size, it is not always possible to return exact output values, but \juanggg{easier to return} \emph{fuzzy} (or approximate) matches. 
For a given input $x$, we obtain an output $\tilde{f}(x)$ that approximates the original function ${f}(x)$. 
First, we use a function $a(x)$ that returns an address. $a(x)$ typically maps a range of inputs $x$ to a single address. For example, $a(x)$ may \juanggg{simply} round down, such that all values in the range $[0, 1)$ map to address $0$. 
Second, we access the \juanggg{lookup} table with $a(x)$ as the address, \juanggg{and} the table returns $l(a(x)) = \tilde{f}(x)$. 

In order to generate the \juanggg{lookup} table, we need a helper function $a^{-1}()$, which is the pseudo-inverse of $a(x)$ in the sense that $x = a(a^{-1}(x)), \forall x$. 
For the previous example, $a^{-1}(0)$ returns one specific value in the range $[0, 1)$. Thus, address $0$ of the lookup table will contain $f(a^{-1}(0))$, i.e., the lookup result for any input in $[0, 1)$. 
The function $a^{-1}()$ determines the spacing between table entries and, thus, the maximum error. For example, in $[0, 1)$ the maximum error will be different if the table contains $f(0)$ or $f(0.5)$. 
A good spacing places more entries where the function's slope is steeper, in order to minimize the error. As a result, the spacing is in proportion to the function's first derivative ($f'(x)$). 
For this reason, the selection of a good $a^{-1}()$ is not always trivial. However, $a^{-1}()$ is only used during table generation, not during table lookups, which allows us to improve the \juanggg{accuracy} (minimize the maximum error) without affecting performance. 

We can use \emph{interpolation} to further improve the \juanggg{accuracy}. In this case, we query two lookup table entries, e.g., $l(a(x))$ and $l(a(x) + 1)$. 
Then, we approximate $f(x)$ as $\tilde{f}(x) = l(a(x)) + l(a(x) + 1) - l(a(x)) \cdot \Delta$, where $\Delta$ represents where the input resides between the two lookup table entries. For example, $\Delta = 0.5$ means that the input $x$ is exactly in between the two entries. 
In the general case, $\Delta = \frac{x - a^{-1}(a(x))}{a^{-1}(a(x) + 1) - a^{-1}(a(x))}$, which needs to be calculated for each input $x$. This calculation can be greatly simplified for our target functions, as we show in Section~\ref{sec:lib-m-lut}. 
Regarding the spacing between entries ($a^{-1}()$), 
\juanggg{the error grows when a function's slope changes quickly (i.e., when the \emph{rate of change} of the function's first derivative is high)}. 
Thus, a desirable spacing should follow the function's second derivative ($f''(x)$).

\subsubsection{Range Extensions}
\label{sec:range}
Both CORDIC and lookup tables support limited ranges of inputs. 
For some transcendental functions, it is possible to extend the range by performing a conversion that depends on the function itself. 
For example, trigonometric functions (e.g., sine, cosine) take advantage of the fact that the output repeats every $2\pi$. 
For other functions, such as square root, logarithm, and exponentiation, we can separate exponent and mantissa. For example, to calculate the logarithm of $x = 2^{exponent(x)} \cdot mantissa(x)$, we convert $log(x) = log(2) \cdot exponent(x) + log(mantissa(x))$.

\section{\libname: Implementation}
\label{sec:implementation}
Given their relatively simple hardware, general-purpose PIM cores do not support complex operations \juanggg{(e.g.,~\cite{gomezluna2022ieeeaccess, upmem-sdk, lee2021hardware, kwon202125})}, such as transcendental (e.g., trigonometric, logarithm, exponentiation) and other hard-to-calculate functions (e.g., square root). 
To fill this gap, we present \emph{\libname}, a library of transcendental and other hard-to-calculate functions for general-purpose PIM cores. 
\libname~leverages the \juanggg{implementation} methods introduced in Section~\ref{sec:transcendental} (\juanggg{e.g.,} CORDIC, lookup tables), and variations and combinations of them. 
Table~\ref{tab:methods} shows \libname's \juanggg{implementation} methods and supported functions.\footnote{\juan{We provide all \juanggg{implementation} methods for sine on the UPMEM PIM architecture~\cite{upmem}. Based on our preliminary analysis of these methods, we also provide the most suitable methods for each of the other supported functions. Future work can extend \libname~with new supported functions.}}

\begin{table}[h]
\caption{\libname's \juanggg{Implementation} Methods and \juanggg{Supported Functions}} 
\label{tab:methods}
\resizebox{\columnwidth}{!}{










\begin{tabular}{l|cccccccccc|}
\cline{2-11}
                                                               & \multicolumn{10}{c|}{\textbf{\juanggg{Supported Functions}}}                                                                                                                                                                                                                                                                                                    \\ \hline
\multicolumn{1}{|l||}{\textbf{\juanggg{Implementation} Method}} & \multicolumn{1}{c|}{\textbf{sin}} & \multicolumn{1}{c|}{\textbf{cos}} & \multicolumn{1}{c|}{\textbf{tan}} & \multicolumn{1}{c|}{\textbf{sinh}} & \multicolumn{1}{c|}{\textbf{cosh}} & \multicolumn{1}{c|}{\textbf{tanh}}       & \multicolumn{1}{c|}{\textbf{exp}}        & \multicolumn{1}{c|}{\textbf{log}}        & \multicolumn{1}{c|}{\textbf{sqrt}}       & \juanggg{\textbf{GELU}} \\ \hline \hline
\multicolumn{1}{|l||}{CORDIC}                                   & \multicolumn{1}{c|}{\checkmark}   & \multicolumn{1}{c|}{\checkmark}   & \multicolumn{1}{c|}{\checkmark}   & \multicolumn{1}{c|}{\checkmark}    & \multicolumn{1}{c|}{\checkmark}    & \multicolumn{1}{c|}{\checkmark} & \multicolumn{1}{c|}{\checkmark} & \multicolumn{1}{c|}{\checkmark} & \multicolumn{1}{c|}{\checkmark} &                \\ \hline
\multicolumn{1}{|l||}{M-LUT}                                    & \multicolumn{1}{c|}{\checkmark}   & \multicolumn{1}{c|}{\checkmark}   & \multicolumn{1}{c|}{\checkmark}   & \multicolumn{1}{c|}{}              & \multicolumn{1}{c|}{}              & \multicolumn{1}{c|}{}           & \multicolumn{1}{c|}{\checkmark} & \multicolumn{1}{c|}{\checkmark} & \multicolumn{1}{c|}{\checkmark} &                \\ \hline
\multicolumn{1}{|l||}{M-LUT+Interpolation}                      & \multicolumn{1}{c|}{\checkmark}   & \multicolumn{1}{c|}{\checkmark}   & \multicolumn{1}{c|}{\checkmark}   & \multicolumn{1}{c|}{}              & \multicolumn{1}{c|}{}              & \multicolumn{1}{c|}{}           & \multicolumn{1}{c|}{\checkmark} & \multicolumn{1}{c|}{\checkmark} & \multicolumn{1}{c|}{\checkmark} &                \\ \hline
\multicolumn{1}{|l||}{L-LUT}                                    & \multicolumn{1}{c|}{\checkmark}   & \multicolumn{1}{c|}{\checkmark}   & \multicolumn{1}{c|}{\checkmark}   & \multicolumn{1}{c|}{}              & \multicolumn{1}{c|}{}              & \multicolumn{1}{c|}{}           & \multicolumn{1}{c|}{\checkmark} & \multicolumn{1}{c|}{\checkmark} & \multicolumn{1}{c|}{\checkmark} &                \\ \hline
\multicolumn{1}{|l||}{L-LUT+Interpolation}                      & \multicolumn{1}{c|}{\checkmark}   & \multicolumn{1}{c|}{\checkmark}   & \multicolumn{1}{c|}{\checkmark}   & \multicolumn{1}{c|}{}              & \multicolumn{1}{c|}{}              & \multicolumn{1}{c|}{}           & \multicolumn{1}{c|}{\checkmark} & \multicolumn{1}{c|}{\checkmark} & \multicolumn{1}{c|}{\checkmark} &                \\ \hline
\multicolumn{1}{|l||}{D-LUT+Interpolation}                      & \multicolumn{1}{c|}{\checkmark}   & \multicolumn{1}{c|}{}             & \multicolumn{1}{c|}{}             & \multicolumn{1}{c|}{}              & \multicolumn{1}{c|}{}              & \multicolumn{1}{c|}{\checkmark} & \multicolumn{1}{c|}{}           & \multicolumn{1}{c|}{}           & \multicolumn{1}{c|}{}           & \checkmark     \\ \hline
\multicolumn{1}{|l||}{DL-LUT+Interpolation}                     & \multicolumn{1}{c|}{\checkmark}   & \multicolumn{1}{c|}{}             & \multicolumn{1}{c|}{}             & \multicolumn{1}{c|}{}              & \multicolumn{1}{c|}{}              & \multicolumn{1}{c|}{\checkmark} & \multicolumn{1}{c|}{}           & \multicolumn{1}{c|}{}           & \multicolumn{1}{c|}{}           & \checkmark     \\ \hline
\multicolumn{1}{|l||}{CORDIC+LUT}                               & \multicolumn{1}{c|}{\checkmark}   & \multicolumn{1}{c|}{\checkmark}   & \multicolumn{1}{c|}{\checkmark}   & \multicolumn{1}{c|}{\checkmark}    & \multicolumn{1}{c|}{\checkmark}    & \multicolumn{1}{c|}{\checkmark} & \multicolumn{1}{c|}{\checkmark} & \multicolumn{1}{c|}{}           & \multicolumn{1}{c|}{}           &                \\ \hline
\end{tabular}
}
\end{table}

\juanggr{\libname~provides files~\cite{transpimlib2023repo} to be included, using the \texttt{\#include} directive, in the host CPU code (for the necessary setup, e.g., loading lookup tables) and the PIM core code for the different implementation methods. 
The APIs are simple and intuitive. For example, for the sine function: \texttt{float sinf (float x);}}

\subsection{CORDIC-based Implementations}
\label{sec:lib-cordic}
\libname~contains CORDIC implementations of trigonometric (sin, cos, tan) and hyperbolic (sinh, cosh, tanh) functions, exponentiation, logarithm, and square root. 

\begin{figure}[h]
\centering
\includegraphics[width=0.7\linewidth]{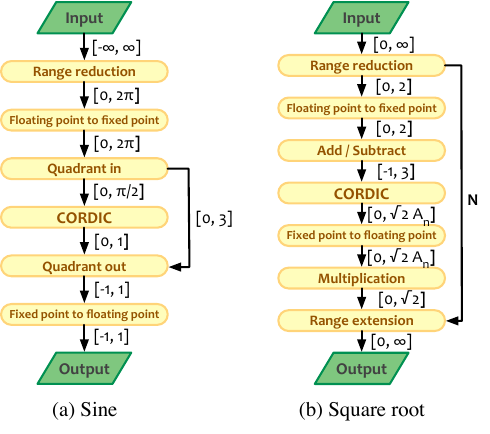}
\caption{\libname's \juanggg{CORDIC-based implementations of sine and square root functions}.}
\label{fig:cordic-blocks}
\end{figure}

We illustrate their implementation \juanggg{using} the sine function as an example. 
As Figure~\ref{fig:cordic-blocks}(a) shows, the calculation takes six steps. 
First, we translate the input value (angle) to the range $0$ to $2\pi$. 
Second, if the input is a floating-point value, we convert it to fixed-point \juanggg{format}.\footnote{CORDIC can operate with fixed-point arithmetic~\cite{cordicprecision}, which is natively supported by the UPMEM PIM architecture~\cite{gomezluna2022ieeeaccess}.} 
Our fixed-point format uses 28 bits for the fractional part, 3 bits for the integer part (enough to represent up to $2\pi$), and 1 sign bit. 
Third, the range is further reduced to \juanggg{the range from} $0$ to $\frac{\pi}{2}$ (the quadrant, \juanggg{i.e., $0$ to $\frac{\pi}{2}$, $\frac{\pi}{2}$ to $\pi$, $\pi$ to $\frac{3\pi}{2}$, $\frac{3\pi}{2}$ to $2\pi$,} is also saved to not lose information). 
Fourth, the CORDIC algorithm iterates until it obtains the sine of the input angle. 
Fifth, we adjust the sine value based on the quadrant of the input angle. 
Sixth, we convert the output value back to floating point \juanggg{format}. 

Other functions may require additional steps. 
For example, \juanggg{our} square root \juanggg{implementation} (Figure~\ref{fig:cordic-blocks}(b)) \juanggg{uses} a final range extension \juanggg{step}. 

\libname~does \emph{not} implement multiplication and division (see Table~\ref{tab:cordicmethods}), which are not natively supported by the UPMEM PIM architecture, because the emulations of the runtime library~\cite{gomezluna2022ieeeaccess} have similar complexity as \juanggg{potential} CORDIC-based implementations and no range limitations.

\subsection{Lookup Table-based Implementations}
\label{sec:lib-table}
\libname~implements several lookup table-based methods that differ in their address generation functions ($a()$ and $a^{-1}()$). 
These methods provide different tradeoffs, as we explain next. 

\subsubsection{Multiplication-based Fuzzy Lookup Table (M-LUT)}
\label{sec:lib-m-lut}
\begin{figure}[h]
\centering
\includegraphics[width=1.0\linewidth]{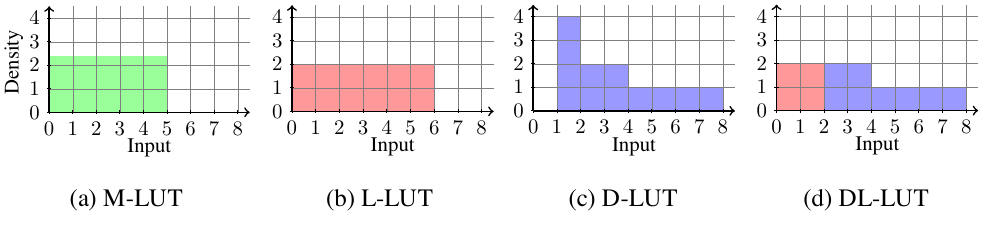}
\caption{Example lookup table density (y axis) for the input range [0, 5] (x axis) and 12 LUT entries \juanggg{for \libname's four different LUT implementations}.}
\label{fig:lut_density}
\end{figure}

This method defines regular spacing between table entries~\cite{mesa2012}. 
We define the address generation function $a(x) = round((x - p) \cdot k)$, where $p$ and $k$ are constants. 
$k$ represents the density of the lookup table, which is the inverse of the spacing. 
$p$ defines what input value $x$ corresponds to address $0$ \juanggg{of the lookup table}. 
For example, to map the interval [0, 5] to a 12-entry M-LUT, we can use $k = \frac{12}{5} = 2.4$ and $p = \frac{5}{2 \cdot 12} = 0.20834$. 
Thus, an input $x = 3$ translates to $a(3) = round((3 - 0.20834) \cdot 2.4) = round(6.7) = 7$. 
Figure~\ref{fig:lut_density}(a) depicts the coverage of a 12-entry M-LUT with density 2.4 for the interval [0, 5].

For the M-LUT, the inverse operation is $a^{-1}(a(x)) = \frac{a(x)}{k} + p$. 
For our previous example, address $7$ of the M-LUT stores the exact value for the input $x = 7 / 2.4 + 0.20834 = 3.125$.

The M-LUT's address generation needs one subtraction, one multiplication, and one step of rounding or truncation. 
For interpolated M-LUTs, we use the address generation function ${a}(x) = floor((x - p) \cdot k)$ to get the next smaller lookup table address. 
The calculation of $\Delta$ simplifies to $\Delta = floor((x - p) \cdot k) - (x - p) \cdot k$. 
As a result, with respect to the M-LUT, the interpolated M-LUT needs one extra \juanggg{lookup} table query, one extra multiplication (to compute $\tilde{f}(x)$), and one extra subtraction (to compute $\Delta$).

\subsubsection{LDEXP-based Fuzzy Lookup Table (L-LUT)}
\label{sec:lib-l-lut}
Multiplication is generally expensive~\cite{gomezluna2022ieeeaccess, gomezluna2021benchmarking}, but we can make it cheaper if we multiply by $2^n$. 
We use the function \texttt{ldexp(arg, exp)}, which is common in standard math libraries~\cite{ldexp}, to perform the operation $arg \cdot 2^{exp}$. 
This function is not available among UPMEM library functions, but we implemented it in accordance with the C99 standard\juanggg{~\cite{C1999iso}}. 

For the L-LUT, we \juanggg{define} the address generation function \juanggg{as} $a(x) = round((x - p) \cdot 2^n)$, which loses some freedom to design lookup tables \juanggg{(the density $k$ must be a power-of-two), but avoids costly multiplication~\cite{gomezluna2022ieeeaccess, gomezluna2021benchmarking}}. 
For the example of a 12-entry lookup table for the interval [0, 5], we can no longer have density $k = 2.4$, but $k = 2$ (i.e., a power of $2$). 
This results in an L-LUT of lower precision than the M-LUT, but expands the range to the interval [0, 6] (Figure~\ref{fig:lut_density}(b)).


\subsubsection{Direct Float Conversion-based Fuzzy Lookup Table (D-LUT)}
\label{sec:lib-d-lut}
As we mention in Section~\ref{sec:fuzzy}, a good spacing depends on the approximated function to minimize the error. 
This may need non-linear address generation functions, which are generally more computationally expensive than multiplications. 
We circumvent this issue by exploiting the natural non-linearity of the floating-point format. 
We propose an address generation function that uses (1) the last $n$ bits of the exponent, and (2) $p$ bits of the mantissa. 
This results in a piece-wise linear density with $2^n$ steps of $2^p$ addresses each. The density of each step is \juanggg{inversely proportional} to the input value, i.e., high density for small inputs and low density for large inputs. 
For example, for our 12-entry table, we can use $n = 2$ (thus, exponents $2^0$, $2^1$, $2^2$) and $p = 2$ (thus, 4 entries per exponent). 
As Figure~\ref{fig:lut_density}(c) shows, the resulting density is 4 in [1, 2), 2 in [2, 4), and 1 in [4, 8). 

The limitation of the D-LUT is that there are no LUT entries between the smallest exponent (e.g., $2^0$) and $0$, which may cause large inaccuracy for LUT queries near $0$. 
To deal with this issue, we propose a combined L-LUT + D-LUT method \juanggg{called \emph{DL-LUT}} (Section~\ref{sec:lib-dl-lut}).

\subsection{Combined Implementations}
\label{sec:combined}
In addition to the previous implementations, \libname~combines pairs of \juanggg{implementation} methods to leverage the strengths of \juanggg{two different implementation methods}. 

\subsubsection{Direct Float Conversion + LDEXP-based Fuzzy Lookup Tables (DL-LUT)} 
\label{sec:lib-dl-lut}
This combination solves the limitation of D-LUT (i.e., no entries between the smallest exponent and $0$) by combining a D-LUT with an L-LUT. 
The DL-LUT uses (1) an L-LUT between $0$ and the smallest exponent, and (2) a D-LUT for larger inputs, providing a density pattern as depicted in Figure~\ref{fig:lut_density}(d).

\subsubsection{CORDIC + LDEXP-based Fuzzy Lookup Table (CORDIC+LUT)} 
\label{sec:lib-cl-lut}
Prior work~\cite{arbaugh2004tablecordic} proposes to replace the first few iterations of CORDIC with a LUT (while still updating $\theta_i$). 
This provides a flexible tradeoff between computing cost, table size, and precision, within the bounds of pure CORDIC and pure LUT approaches (\libname~uses L-LUT for this combined method). 

\section{Evaluation}
\label{sec:evaluation}

This section presents our evaluation of \libname~on a real-world PIM system. 
Section~\ref{sec:methodology} introduces our evaluation methodology. 
Section~\ref{sec:microbenchmarks} presents a microbenchmark-based analysis of the different calculation methods used in \libname. 
Section~\ref{sec:benchmarks} presents the evaluation of \libname~for three real workloads on the PIM system and the comparison to their multithreaded CPU implementations \juanggg{of the three real methods}. 

\subsection{Methodology}
\label{sec:methodology}
We evaluate \libname~on a real-world system with the UPMEM PIM architecture. 
The system consists of a host CPU (2-socket Intel Xeon with 32 cores at 2.10 GHz), standard main memory (128 GB), and 20 UPMEM PIM DIMMs (159 GB and 2545 PIM cores at 350 MHz)\juanggg{~\cite{upmem-server}}.

\subsubsection{Microbenchmarks}
\label{sec:method_ubenchmarks}
In Section~\ref{sec:microbenchmarks}, we evaluate \libname's CORDIC, LUT-based, and combined implementations on a single PIM core \juanggg{(DPU in UPMEM terminology)} using microbenchmarks, in order to compare them in terms of (1) performance, (2) \juangg{accuracy}, (3) memory consumption, and (4) setup time. 
Our microbenchmarks compute \juanggg{all our implemented} transcendental functions (\juanggg{i.e.,} all versions of all functions in Table~\ref{tab:methods}) for the elements of an array (of $2^{16}$ floating-point values with random uniform distribution) that resides in a DRAM bank (MRAM in UPMEM terminology). The PIM core moves chunks of the array into the scratchpad memory (WRAM in UPMEM terminology) and operates on each element. 

For performance comparison, we measure total execution cycles using a hardware counter~\cite{upmem-guide}. 

For \juangg{accuracy} comparison, we compare to the output of the host CPU, computed with the standard math library. We obtain the root-mean-square absolute error (RMSE), which we analyze in Section~\ref{sec:microbenchmarks}. Maximum absolute error and error in terms of units of last place (ULP)~\cite{goldberg91} show very similar trends to the RMSE. 

For memory consumption, we account for all tables and variables that we allocate in the DRAM bank of a PIM core. 

For setup time, we include the generation of any tables or variables on the host CPU, and their transfers to the DRAM bank of a PIM core.


\subsubsection{Benchmarks}
\label{sec:method_benchmarks}
In Section~\ref{sec:benchmarks}, we implement three full workloads that use transcendental functions \juanggg{(Blackscholes, Sigmoid, and Softmax)} on the UPMEM PIM architecture, and compare them to their CPU-only versions.

Blackscholes~\cite{macbeth1979empirical, bienia2008parsec} calculates the prices for a portfolio of options with a partial differential equation. This benchmark uses several functions that benefit from \libname: exponentiation, logarithm, square root, and cumulative normal distribution function (CNDF). 
The original benchmark implements CNDF using polynomial approximation. We also implement CNDF using \libname's LUT-based methods. In our experiments, we use an input vector of 10M elements.

Sigmoid~\cite{han1995influence} is a bounded differentiable function whose derivative is always positive. A general equation of the Sigmoid function with a scalar input $x$ is defined as $S(x)=\frac{1}{1+e^{-x}}$. Sigmoid is commonly used in logistic regression~\cite{hosmer2013applied} to compute the probability of an output event. It is also frequently used as an activation function in neural networks. Our Sigmoid benchmark takes an input vector \juan{of 30M elements} and computes the Sigmoid output of each input element. 

Softmax~\cite{alpaydin2020} is a function that turns a vector of $K$ real values into a vector of $K$ real values that sum to 1 ($\sigma (\mathbf {z} )_{j}={\frac {e^{z_{j}}}{\sum _{k=1}^{K}e^{z_{k}}}}$). Softmax is frequently used as the last layer of neural networks and in reinforcement learning.\footnote{At the time of writing, there are no public implementations of neural networks \juanggg{(e.g., convolutional neural networks)} or reinforcement learning for the UPMEM PIM architecture where we can test \libname-based Softmax. Such implementations are out of the scope of this work but \libname~can be a useful resource for them.} In our experiments, the input vector contains 
\juan{30M} values. 


On the UPMEM PIM architecture, we implement PIM baselines of \juanggg{these} three workloads \juanggg{that} do \emph{not} use \libname~functions. \juanggg{These baseline PIM} implementations are \juanggg{instead} based on polynomial approximation~\cite{transc_taylor_minmax, transc_taylor}. 
For the PIM implementations that use \libname, we use interpolated M-LUT and L-LUT methods for both Blackscholes, \juanggg{Sigmoid}, and Softmax. 
Additionally, we implement a version of Blackscholes that operates on fixed-point values, \juanggrr{and versions of Sigmoid and Softmax that use CORDIC+LUT}.

\subsection{Microbenchmarks Results}
\label{sec:microbenchmarks}
We use microbenchmarks to evaluate all \juanggg{implementation} methods and \juanggg{supported} functions in Table~\ref{tab:methods}. 
In this section, we analyze the different \juanggg{implementation} methods for the sine function, as a representative function. 
\juanggg{We evaluate floating-point versions of all implementation methods, and also} fixed-point versions of L-LUT methods. 
Our observations and takeaways are applicable to the \juanggg{other functions as well}. 

We run experiments for all CORDIC, LUT-based, and combined methods where we tune the methods (e.g., number of iterations in CORDIC-based methods, LUT size in LUT-based methods) to obtain 
\juangg{an accuracy} range, i.e., different root-mean-square absolute errors  between $10^{-4}$ and $10^{-9}$. 
The desired \juangg{accuracy} impacts the execution time (on the PIM side), the setup time (on the host CPU side), and the memory consumption (on the PIM side). Hence, we analyze (1) performance (\juanggg{execution cycles on the} PIM core), (2) setup time (seconds), and (3) memory consumption (bytes) as a function of the root-mean-square absolute error (RMSE). 

\subsubsection{\juanggg{Execution Cycles}}
\begin{figure}[h]
\vspace{-2mm}
\centering
\includegraphics[width=1.0\linewidth]{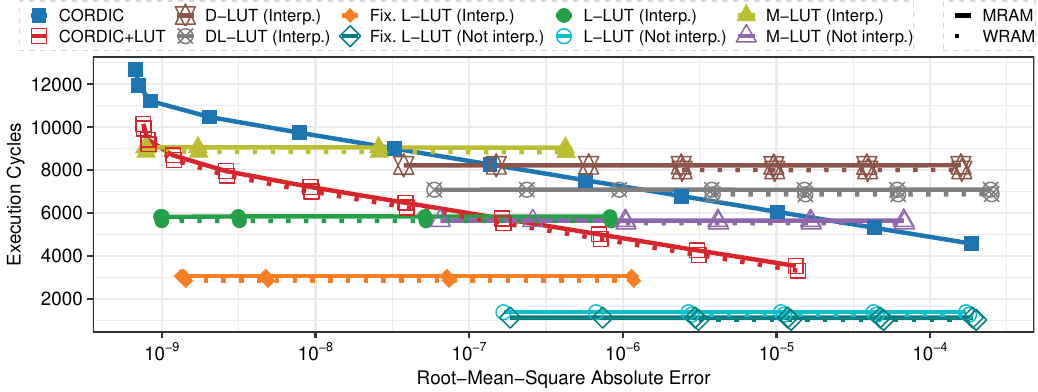}
\vspace{-20pt}
\caption{Execution cycles (y axis, linear scale) per input element on one PIM core \juanggg{as a function of} root-mean-square absolute error \juanggg{values} (x axis, logarithmic scale) for \juanggg{different} \libname~implementations of the sine function.}
\label{fig:cycles}
\end{figure}

Figure~\ref{fig:cycles} shows the execution cycles (per input element) on a PIM core for the different \libname~methods that implement the sine function \juanggg{as a function of the accuracy provided by the method}. 
LUT-based versions place the LUT in \juanggg{either} the PIM core's DRAM bank (MRAM, solid line) or the scratchpad (WRAM, dashed line). 
We make five observations. 


First, each LUT-based method consumes the same \juanggg{number} of cycles for any RMSE, because the number of LUT accesses and operations is independent of the LUT size. 
LUT-based methods with the same \juanggg{number} of floating-point multiplications take similar \juanggg{number} of execution cycles,  
\juanggg{i.e., the number of floating-point multiplications determines the number of execution cycles. 
The slowest of the LUT-based methods is the interpolated M-LUT method (yellow line), which executes two floating-point multiplications per input element}. 
Non-interpolated M-LUT (purple) and interpolated L-LUT (green) execute one multiplication, while non-interpolated L-LUT (\juan{cyan}) needs no multiplications. 
As a result, L-LUT methods (with our custom LDEXP operation, Section~\ref{sec:lib-l-lut}) outperform M-LUT methods for the entire \juangg{accuracy} range. 
Interpolated L-LUT reduces the cycle count by ${\sim}50\%$ over interpolated M-LUT, and non-interpolated L-LUT reduces the count by ${\sim}80\%$ over non-interpolated M-LUT. 
\juanggg{The fixed-point version of the non-interpolated L-LUT method (teal)} does not improve the performance over its floating-point \juanggg{counterpart (none of them use multiplications)}. 
However, \juanggg{the fixed-point version of} the interpolated L-LUT (orange) doubles the performance of the floating-point version \juanggg{of the interpolated L-LUT (green)} on the UPMEM PIM architecture, where floating-point multiplications are significantly more costly than fixed-point multiplications~\cite{gomezluna2022ieeeaccess, gomezluna2021benchmarking}.

Second, CORDIC-based methods \juanggg{take} more execution cycles \juanggg{to provide} higher \juangg{accuracy}, because \juangg{accuracy} increases with each iteration of the CORDIC algorithm. 
CORDIC+LUT runs faster than pure CORDIC, as it \juanggg{replaces the initial iterations of the CORDIC algorithm} with an L-LUT query. 

Third, L-LUT methods obtain the best tradeoff in terms of performance and \juangg{accuracy}. \juanggg{Note that curves that are closer to the bottom-left corner of the plot (e.g., interpolated floating-point L-LUT, interpolated fixed-point L-LUT) are better}. 
\juanggg{Interpolated floating-point and fixed-point L-LUT methods are only 2-3 times more costly than a single-precision floating-point and an integer multiplication, respectively}. 
As a result, L-LUT methods have an inherent advantage on PIM architectures over other methods, such as Taylor approximation~\cite{transc_taylor, transc_taylor_minmax}, where one floating-point multiplication is needed for each bit of precision (e.g., ${\sim}28$ \juanggg{multiplications} for RMSE=$10^{-9}$).

Fourth, there is no significant performance difference between allocating and accessing LUTs in the DRAM bank (MRAM, solid line) or in the scratchpad (WRAM, dashed line), \juan{and this observation holds for any number of PIM threads running on the PIM core}. 
\juan{However, the scratchpad is significantly smaller than the DRAM bank. Thus, the \juangg{accuracy} is limited by the maximum possible LUT size (most noticeably for non-interpolated methods, e.g., in Figure~\ref{fig:cycles}, non-interpolated fixed-point L-LUT results in RMSE={${\sim}10^{-6}$} in WRAM (teal, dashed) but {${\sim}10^{-7}$} in MRAM (teal, solid)).} 
As such, placing LUTs in MRAM can be a good choice to save WRAM \juanggg{space} for input/output operands. 

Fifth, at around RMSE=$10^{-9}$, further increasing the LUT size or the number of CORDIC iterations does \emph{not} provide further \juangg{accuracy} increase. 
In our experiments, we also measure \juanggg{the} maximum absolute error \juanggg{to be} around $10^{-7}$ (not shown in Figure~\ref{fig:cycles}). 
Intuitively, this is due to the precision of floating-point values being limited to $4 \cdot 2^{-24}$ ($2.38 \cdot 10^{-7}$) for inputs in the range $[4, 8)$ (inputs values are in the range $[0, 2\pi]$). 
For fixed-point values, we use \juanggg{a} 28-bit fractional part, \juanggg{such that} the precision of $2^{-28}$ ($3.7 \cdot 10^{-9}$) is sufficient to match the \juangg{accuracy} \juanggg{provided by} floating-point values. 
This makes fixed-point L-LUT methods a good choice for general-purpose PIM architectures such as UPMEM PIM, where floating-point computation is not natively supported.

\takeaway{\juanggr{Interpolated L-LUT methods (lookup table with LDEXP operation) offer the best tradeoff in terms of performance and accuracy.}}

\subsubsection{\juanggg{Setup Time in Host CPU}}
\begin{figure}[h]
\vspace{-2mm}
\centering
\includegraphics[width=1.0\linewidth]{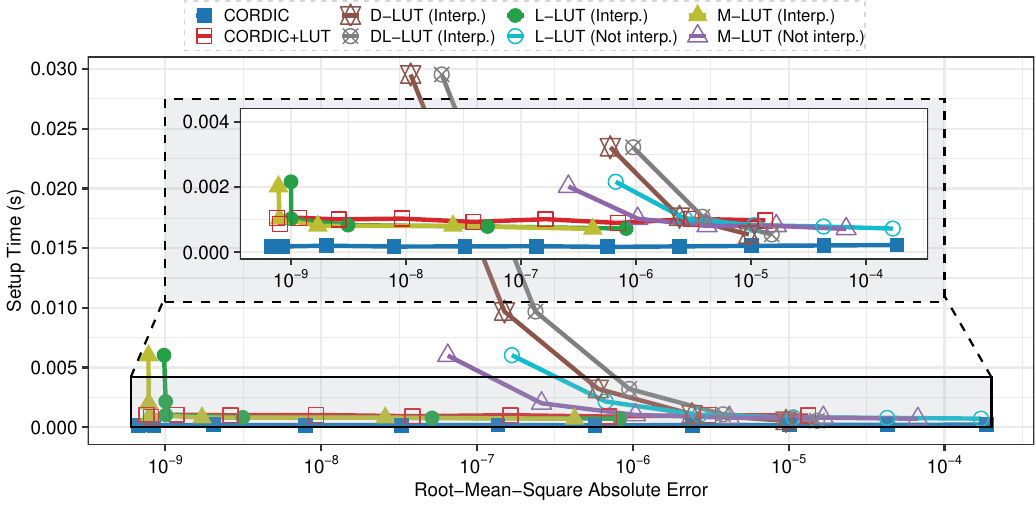}
\vspace{-20pt}
\caption{Setup time in seconds (y axis, linear scale) on the host CPU \juanggg{as a function of} root-mean-square absolute error \juanggg{values} (x axis, logarithmic scale) for \juanggg{different} \libname~implementations of the sine function.}
\label{fig:setup}
\end{figure}

Figure~\ref{fig:setup} shows the setup time on the host CPU for each \juanggg{implementation} method \juanggg{as a function of the accuracy provided by the method}. 
In combination with the execution cycles on the PIM core, the setup time gives us a more complete understanding about when to use CORDIC-based or LUT-based methods. 
We make two observations. 

First, CORDIC-based methods have flat setup times while LUT-based methods have setup times that increase with the table size. 
We observe that pure CORDIC \juanggg{implementations} can provide higher performance \juanggg{(i.e., lower setup time)} than LUT-based methods when the total number of transcendental operations in a workload is low. 
For example, we can compare L-LUT and pure CORDIC at RMSE=$10^{-9}$: (1) CORDIC takes 5380 cycles more than L-LUT on the PIM core, and (2) L-LUT's setup time on the host CPU is ${\sim}5\cdot10^{-4}$ seconds longer than CORDIC's setup time. With a PIM core running at 425 MHz, a PIM kernel would need to execute ${\sim}40$ sine operations for the L-LUT sine to amortize the setup time with respect to the CORDIC-based sine. 
Thus, CORDIC appears to be preferable for kernels computing just a few transcendental functions \juanggg{(e.g., less than 40 sine operations in the previous example)}. 

Second, CORDIC+LUT has higher setup times than pure CORDIC due to the use of a LUT for the initial iterations. 
However, CORDIC+LUT's setup times are \juanggg{mostly} flat \juanggg{due to the use of CORDIC for later iterations}. 
\juanggg{Compared to CORDIC and L-LUT,} 
CORDIC+LUT \juanggg{is} a good choice for kernels with very few transcendental functions that require high \juangg{accuracy}. 

\takeaway{\juanggr{CORDIC-based methods are preferable when a PIM kernel needs to execute just a few transcendental functions (e.g., less than 40 sine operations in a kernel running on the UPMEM PIM architecture) due to their low setup time in the host CPU.}}

\subsubsection{\juanggg{Memory Consumption}}
\begin{figure}[h]
\centering
\includegraphics[width=1.0\linewidth]{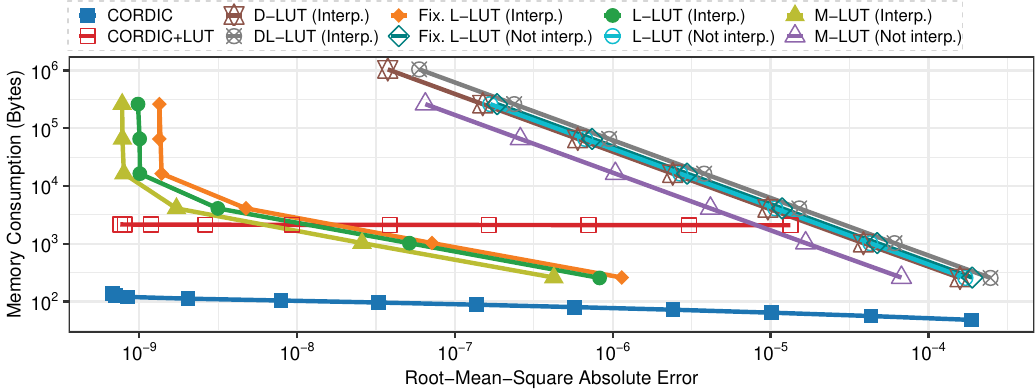}
\vspace{-20pt}
\caption{Memory consumption in bytes (y axis, logarithmic scale) per PIM core \juanggg{as a function of} root-mean-square absolute error \juanggg{values} (x axis, logarithmic scale) for \juanggg{different} \libname~implementations of the sine function.}
\label{fig:bytes}
\end{figure}

Figure~\ref{fig:bytes} shows the memory consumption (in bytes) per PIM core of all LUT- and CORDIC-based \juanggg{implementation} methods \juanggg{as a function of root-mean-square absolute error values}. 
We make several observations. 


First, \juangg{accuracy} of non-interpolated LUT-based methods is limited by the amount of available memory (DRAM bank or scratchpad). 
Thus, they are only recommended when fast calculation is needed but lower \juangg{accuracy} is acceptable. 

Second, CORDIC and CORDIC+LUT methods have the advantage that their memory consumption does \emph{not} grow exponentially. 
Thus, they are recommended for applications \juanggg{that require high accuracy}, where the amount of memory that can be devoted to \libname~is limited. \juanggg{For example, this can happen in applications with large datasets,} where we need to allocate most of the space in the PIM core's DRAM bank to input/output operand arrays. 

Third, interpolation is an effective way of increasing \juangg{accuracy} \juanggg{without increasing} LUT size. \juanggg{Overall,} interpolated L-LUT offers a good tradeoff in terms of \juangg{accuracy}, execution cycles, and memory consumption. 
For example, at the maximum \juangg{accuracy} that non-interpolated LUT-based methods can provide (${\sim}10^{-7}$), interpolated L-LUT needs less memory than CORDIC+LUT and it is significantly faster than pure CORDIC (see Figure~\ref{fig:cycles}). 

\takeaway{\juanggr{Interpolated L-LUT methods offer a good tradeoff in terms of accuracy, execution cycles, and memory consumption. However, CORDIC and CORDIC+LUT methods are recommended for applications that require high accuracy, where the available memory is needed for large datasets (i.e., not available for lookup tables required for the necessary accuracy).}}

\subsubsection{\juanggg{Other Supported Functions}}
The general trends for other functions supported by \libname~are similar to those of the sine function, which we discuss above. 
Some major differences that \juanggg{are worth highlighting} are as follows. 

First, methods for tangent calculation take around 2-3 times more cycles than the same methods for sine. 
This is explained by the fact that tangent needs (1) calculation of sine \emph{and} cosine, and (2) a floating-point division (\juanggg{much costlier than a} floating-point multiplication on UPMEM~\cite{gomezluna2022ieeeaccess, gomezluna2021benchmarking}). 

\begin{figure}[h]
\vspace{-2mm}
\centering
\includegraphics[width=0.75\linewidth]{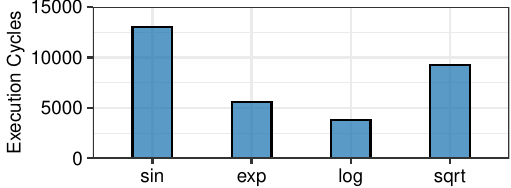}
\vspace{-3mm}
\caption{Execution cycles per input element for range reduction/extension of \libname~implementations of the sine, exponential, logarithm, and square root functions.}
\label{fig:extension}
\end{figure}

Second, some \juanggg{supported} functions may require range reduction and/or range extension (Section~\ref{sec:range}), e.g., sine/cosine, exponentiation, logarithm, square root. 
The cost of these operations largely \juanggg{differs} between functions, because it depends on the specific operations needed for the conversion (\juanggg{e.g.}, mathematical identity that applies to each function, \juanggg{Section~\ref{sec:range}}). 
Figure~\ref{fig:extension} shows the execution \juanggg{cycles} per input element for range reduction/extension in \emph{sin}, \emph{exp}, \emph{log}, and \emph{sqrt}. 
Note that range reduction/extension is only necessary depending on the range of input values. For example, our experiments with the sine function (Figures~\ref{fig:cycles} to~\ref{fig:bytes}) use input values in $[0, 2\pi)$.


Third, functions that do \emph{not} need range reduction/extension are cheaper to calculate. 
This is the case \juanggg{for} activation functions such as \emph{tanh} and GELU \juanggg{(Gaussian Error Linear Unit)~\cite{gelu}}, which \juanggg{are approximately linear in most parts}. 
D-LUT and DL-LUT methods are particularly well-suited for \emph{tanh} and GELU, \juanggg{unlike} sine (Figure~\ref{fig:cycles}). They are ${\sim}2\times$ faster than, e.g., interpolated L-LUT for sine, while providing similar \juangg{accuracy}. 


\takeaway{\juanggr{D-LUT and DL-LUT methods are well-suited for activation functions, such as \emph{tanh} and GELU, which (1) do not require range extension, and (2) are approximately linear in most parts. They are faster than interpolated L-LUT, while providing similar accuracy.}}

\subsection{\juanggg{Real-World Benchmark} Results}
\label{sec:benchmarks}

We implement for the UPMEM PIM architecture \juan{three} full workloads (Section~\ref{sec:method_benchmarks}) that make use of several functions that are supported by \libname. 
We compare them to their single-thread and 32-thread CPU baselines \juanggg{as well as a PIM baseline that uses polynomial approximation}. 

\begin{figure}[ht!]
\centering
\begin{subfigure}{0.8\linewidth}
    \includegraphics[width=\linewidth]{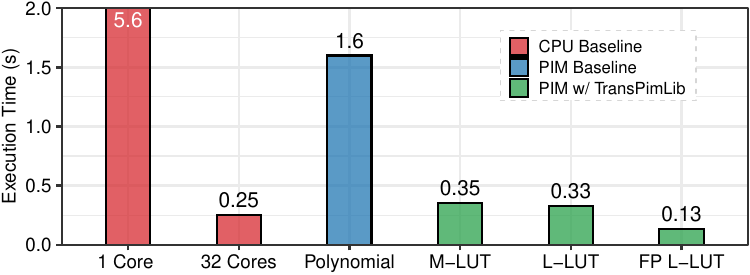}
    \vspace{-15pt}
    \caption{Blackscholes}
    \vspace{5pt}
    \label{fig:first}
\end{subfigure}
\vfill
\begin{subfigure}{0.8\linewidth}
    \includegraphics[width=\linewidth]{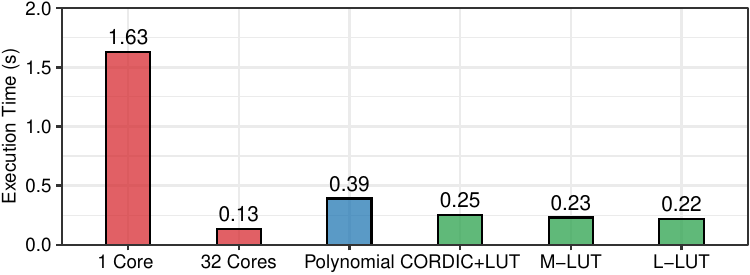}
    \vspace{-15pt}
    \caption{Sigmoid}
    \vspace{5pt}
    \label{fig:second}
\end{subfigure}
\vfill
\begin{subfigure}{0.8\linewidth}
    \includegraphics[width=\linewidth]{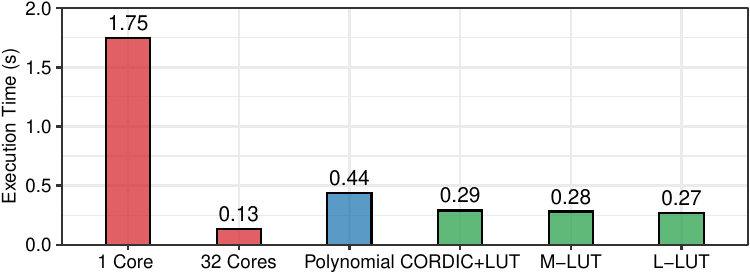}
    \caption{Softmax}
    \vspace{5pt}
    \label{fig:third}
\end{subfigure}
\vspace{-15pt}   
\caption{Execution time (s) of Blackscholes (a), Sigmoid (b), and Softmax (c) implementations on PIM \juanggg{(on 2545 PIM cores running 16 PIM threads each)}, 1 CPU core, and 32 CPU cores. }
\label{fig:benchmarks}
\end{figure}

Figure~\ref{fig:benchmarks} shows the execution time of PIM implementations of Blackscholes, \juan{Sigmoid}, and Softmax \juanggg{(on 2545 PIM cores running 16 PIM threads each)}, and the CPU baselines \juanggg{(on 1 and 32 CPU cores)}. 
We test PIM versions that use (1) polynomial approximation~\cite{transc_taylor, transc_taylor_minmax} (for comparison to \libname's methods), (2) interpolated M-LUT, and (3) interpolated L-LUT. For Blackscholes, we also test a version with interpolated fixed-point L-LUTs. 
\juanggrr{For Sigmoid and Softmax, we test versions with CORDIC+LUT.} 
\juan{We include in the measurements all range reduction/extension costs needed for some functions (Figure~\ref{fig:extension}).} 
We make the following observations from Figure~\ref{fig:benchmarks}. 

First, the PIM versions of Blackscholes that use \libname~reduce the execution time by \juanggrr{$5-12\times$} with respect to the \juanggg{baseline} PIM version with polynomial approximation. 
\juan{The M-LUT and L-LUT versions are, respectively, within \juanggrr{71\% and 75\%} the performance of the 32-thread CPU baseline. 
The fixed-point L-LUT version \juanggg{(on 2545 PIM cores running 16 PIM threads each)}  is \juanggrr{92\%} faster than the 32-thread CPU baseline.}

Second, both Sigmoid and Softmax show similar \juanggg{qualitative} behavior. 
\libname's methods outperform the PIM version with polynomial approximation by \juanggrr{52-77\%}. 
The 32-thread CPU baselines are around $2\times$ faster than \libname's PIM version. 
Nonetheless, these functions are typically used as part of neural networks and machine learning algorithms, which may run on the PIM cores. 
Thus, \libname's methods can reduce data movement from PIM cores to the CPU (Figure~\ref{fig:transcendental}(b)) for applications running on the PIM cores. 
As a result of saving \juanggg{such} PIM-Host and Host-PIM transfers, the execution of transcendental functions in the PIM cores (Figure~\ref{fig:transcendental}(c)) \juanggg{could} be $6-8\times$ faster than the execution in the host CPU (\juanggg{as inferred from} Figure~\ref{fig:transcendental}(b)).

\takeaway{\juanggrr{\libname~can reduce data movement from PIM cores to the CPU (Figure~\ref{fig:transcendental}(b)) for applications running on the PIM cores. 
As a result, the execution of transcendental functions in the PIM cores (Figure~\ref{fig:transcendental}(c)) could be faster than the execution in the host CPU.}}



\section{Related Work}
\label{sec:related}

To our knowledge, \libname~is the first library of transcendental (and other hard-to-calculate) functions for general-purpose PIM systems. 

\subsection{\juanggr{Real Processing-in-Memory Systems}}
PIM has become a reality in the last few years. 
UPMEM~\cite{upmem} was first to release their PIM architecture~\cite{upmem-guide, gomezluna2022ieeeaccess}. Since it is the first publicly-available general-purpose PIM architecture, we have implemented \libname~for it. 

There is a good amount of recent works that analyze the UPMEM PIM architecture and implement important applications for it. 
An experimental characterization of the UPMEM PIM architecture and a benchmark suite is presented in~\cite{gomezluna2022ieeeaccess, gomezluna2021cut}. 
SpMV, an important memory-bound kernel, is extensively explored in~\cite{giannoula2022sigmetrics}. 
The \emph{wavefront algorithm} (\emph{WFA})~\cite{marco2021fast}, which is currently the state-of-the-art gap-affine pairwise alignment algorithm, a key step in genome analysis, is implemented on the UPMEM PIM architecture in~\cite{diab2022high, diab2023aim}. 

Besides UPMEM, there have been several prototypes of real PIM chips developed by major vendors in industry, including Samsung, SK Hynix, and Alibaba, in 2021-2022. 
Samsung introduced HBM-PIM, also known as FIMDRAM,~\cite{kwon202125, lee2021hardware}, an architecture that embeds one floating-point SIMD unit with a reduced instruction set, called Programmable Compute Unit (PCU), next to two DRAM banks in HBM2 layers. This architecture is targeted to accelerate machine learning inference. 
The second prototype from Samsung is AxDIMM~\cite{ke2021near, lee2022improving}. AxDIMM is a DIMM-based solution which places an FPGA fabric in the buffer chip of the DIMM. It has been tested for DLRM recommendation inference~\cite{naumov2019deep, ke2021near} and in-memory databases~\cite{lee2022improving}. 
Another major DRAM vendor, SK Hynix, introduced Accelerator-in-Memory~\cite{lee2022isscc}, a GDDR6-based PIM architecture with specialized units for multiply-and-accumulate and lookup-table-based activation functions for deep learning applications. 
AiM~\cite{lee2022isscc} uses LUTs and interpolation hardware for activation functions. 
A key difference with our work is that AiM requires dedicated hardware for address generation and interpolation. 
Alibaba introduced HB-PNM~\cite{niu2022isscc}, a PNM system with specialized engines for recommendation systems, which is composed of a DRAM die and a logic die vertically integrated via hybrid bonding~\cite{fujun2020stacked}. 
\libname~can be realized for any PIM architecture that supports addition, subtraction, multiplication, and division. As such, future work can implement new versions of \libname's methods for other current and future PIM architectures. 

\juanggr{Though we expect that real PIM systems will continue improving their computing capabilities, integrating processing elements in DRAM technology is challenging and constrains design decisions heavily~\cite{devaux2019, kim1999assessing, keitel2001embedded}. 
For example, DRAM has a lower number of metal layers than CMOS and slower transistors~\cite{devaux2019,yuffe2011,christy2020,singh2017,weber2005current,peng2015design}. 
As a result, manufacturing complex execution units (as those needed for transcendental and other hard-to-calculate functions) using DRAM technology would require the addition of extra (costly) metal layers while resulting in low frequency logic units~\cite{kim1999assessing, keitel2001embedded}, which will hardly be affordable in PIM systems.
Other future PIM systems, e.g., 3D-stacked memories with processing elements in a logic layer~\cite{hmc.spec.1.1, hmc.spec.2.0, jedec.hbm.spec}, can make the integration of complex execution units easier. 
However, their area and thermal budget will still be constrained. For example, in a forward-looking HBM-based PIM system~\cite{kim2021signal}, PIM logic can occupy only ${\sim}28\%$ of the logic layer due to the need for peripheral/control logic. 
Thus, the availability of libraries for complex operations, such as \libname, will likely continue to be necessary.}

\subsection{\juanggr{Acceleration of Transcendental Functions}}
\juanggr{Several works~\cite{symetric_table_addition, exact_lookup_table} aim to improve LUT-based implementation methods (e.g., improving accuracy for a given memory consumption). These works are not specific to PIM systems. Compared to them, \libname~is simpler and requires less calculations before and after lookup table queries.}

\juanggr{Several other works~\cite{memoization1, memoization2, memoization3} study the use of memoization to accelerate expensive transcendental function calls in CPUs. Compared to \libname, these approaches are not self-sufficient and, as such, they need additional mechanisms for the cases where a value that has not been memoized is needed. As a result, if implemented for PIM systems, these approaches could lead to excessive data movement, as shown in Figure~\ref{fig:transcendental}(b).}

\section{Conclusion}
\label{sec:conclusion}
Processing-in-memory (PIM) is a promising trend to alleviate the data movement bottleneck in current computing systems. 
PIM is becoming a reality with the advent of real-world PIM architectures, which place simple processing elements near the memory arrays. 
These architectures support only limited instruction sets, which makes the execution of complex operations challenging. This is the case of transcendental functions and other hard-to-calculate operations (e.g., square root). 

In this work, we present \libname, the first library for PIM systems that provides CORDIC-based and LUT-based methods for trigonometric functions, hyperbolic functions, exponentiation, logarithm, square root, etc. 
We develop an implementation of \libname~for the UPMEM PIM architecture and perform a thorough evaluation of \libname's methods in terms of performance and \juangg{accuracy}, using microbenchmarks and three full workloads (Blackscholes, Softmax, Sigmoid). 

We believe that \libname~methods can be suitable for other \juanggr{current and future PIM architectures lacking native} support for these complex functions. The implementation of these methods for other current and future PIM architectures is subject of future work.

\begin{acks}
\juanggg{This paper appears at ISPASS 2023~\cite{item2023transpimlibispass}. We thank the anonymous reviewers of ISPASS 2023 for feedback.} 
We acknowledge the generous gifts provided by our industrial partners, including ASML, Facebook, Google, Huawei, Intel, Microsoft, and VMware.
We acknowledge support from the Semiconductor Research Corporation, the ETH Future Computing Laboratory, \juanggrr{and the European Union’s Horizon programme for research and innovation under grant agreement No. 101047160, project BioPIM (Processing-in-memory architectures and programming libraries for bioinformatics algorithms)}.
This research was partially supported by ACCESS – AI Chip Center for Emerging Smart Systems, sponsored by InnoHK funding, Hong Kong SAR.


\end{acks}

\balance
\bibliographystyle{plain}
\bibliography{references}

\end{document}